# InnerSelf: Designing Self-Deepfaked Voice for Emotional Well-being

Guang Dai[1], Pinhao Wang[1], Cheng Yao[1,*], Fangtian Ying[1]

1. College of Computer Science and Technology, Zhejiang University, Hangzhou, China.

## Abstract

One's own voice is one of the most frequently heard voices. Studies found that hearing and talking to oneself have positive psychological effects. However, the design and implementation of self-voice for emotional regulation in HCI have yet to be explored. In this paper, we introduce InnerSelf, an innovative voice system based on speech synthesis technologies and the Large Language Model. It allows users to engage in supportive and empathic dialogue with their deepfake voice. By manipulating positive self-talk, our system aims to promote self-disclosure and regulation, reshaping negative thoughts and improving emotional well-being.

## 1. Introduction

Hearing one's own voice is an uncommon experience that has significant psychological implications. Neuroscience research indicates that hearing one's own voice activates specific neural pathways associated with self-awareness and emotional processing [1]. In language learning, incorporating the learner's own voice has been proven its advantages in word recognition and memorization [2], [3], improved pronunciation [4]. Regarding mental health, self-directed speech, when recorded and played back, can help individuals externalize and reframe negative thoughts, a common technique used in cognitive behavioral therapy [5]. Similarly, self-tracking technologies leverage voice as a natural and low-burden input method to support personal data capture. Users can record and send voice messages via smartphones to report daily activities, emotions, and other personal states [6]. In the field of Human-Computer Interaction (HCI), the effects of self-voice have been shown to enhance user attention and goal-directed behavior [7]. When applied to voice alarms as reminders for daily tasks, self-voice alarms significantly improved users' alertness, task completion rates, and the frequency of repeating tasks, thereby supporting the achievement of daily goals. Unlike these previous studies that primarily used pre-recorded voices, it is now possible to create one's own voice using AI-based speech synthesis technology. Individuals can easily clone or replicate their own or others' voices for deepfake voices. One prospective study [8] explored user experience with self-deepfake voices in hedonic and pragmatic contexts, compared to real human voices and AI-generated voices. The results found that self-deepfake voices caused higher emotional responses and a sense of closeness. However, the potential of self-voice interfaces in emotional regulation, its sensory experience and interactivity has not been extensively explored, particularly in the growing demand for digital health solutions and emotionally intelligent agents.

With advances in the Natural Language Processing capabilities of Large Language models, conversational agents and chatbots provide automation benefits and more naturalistic interactions [9]. Our study focuses on emotional regulation with voice assistant technology. Among them, voice interfaces for self-disclosure offer new avenues for mental health research and practice in HCI. For instance, Lee et al. studies [10], [11] introduced a chatbot designed to facilitate deep self-disclosure. Users interacted with the chatbot for journaling, small talk, and answering sensitive questions. The study found that chatbots effectively promoted users' deep self-disclosure on sensitive topics and enhanced their trust and intimacy with the chatbot.

In this paper, our work proposes a self-voice interactive system based on recent achievements in speech synthesis technology and Natural Language Processing. This research aims to assist individuals, through positive self-talk to reshape their negative thoughts and improve their emotional well-being. The contributions of this paper are: First, we identify the effects of hearing and talking to ourselves in terms of emotional regulation, providing a novel approach to voice interfaces for the HCI community. Second, we present the InnerSelf system and discuss its potential applications in immediate managing of emotional reactions, long-term changes in thinking patterns, and providing effective psychological assistance.

## 2. Design of InnerSelf

### 2.1. Design Principles

One's own voice is one of the most important and most frequently heard voices. Studies in related fields found that hearing and talking to oneself has profound effects on affective, cognitive, and behavioral processes. This section explores these findings and how they guided the design of InnerSelf.

Self-related stimuli such as the sound of one's own voice can elicit high levels of emotional arousal, resulting in attentional bias [12], [13]. This means that audiences are more likely to allocate attentional resources to information presented in their own voice than to neutral stimuli. Moreover, individuals exhibit better memory retention for self-related information compared to unrelated information, a phenomenon known as the self-reference effect (SRE) [14]. As highlighted by researchers like Argembeau et al. [15], SRE also varies according to the affective valence of the stimuli. In particular, positive trait information is typically better recalled than negative trait information when encoded in reference to the self. By integrating these effects with HCI methods, it is possible to develop personalized solutions for self-regulation, thus setting the stage for the explorations presented in this paper.

However, listening to one's own voice often induces negative reactions, a phenomenon known as voice confrontation [16]. This cognitive dissonance arises mainly because the sound we hear while speaking is transmitted directly through our bones to our ears. The recorded voice reaches our ears through the air, making it sound lower and unfamiliar, which can cause discomfort and awkward feelings. In contrast, some studies [17] found that most people perceive their own voices as more attractive than others. As the identification of one's voice significantly impacts its evaluation, we suggest the use of bone conduction headphones, which can help to better differentiate between one's own voice and that of others, improving the listener's acoustic experience [18].

In addition, our study aims to manipulate positive self-talk through user interacting with their deepfake self's voices. Self-talk refers to verbalizations addressed to the self. Positive self-talk scripts are the specific language and content of an individual's internal self-communication, consisting of a series of positive, realistic, and constructive sentences or phrases that help the individual to maintain a positive state of mind and behavior in the face of adversity or challenge. Commonly, contents include affirmations (e.g., *"I am capable and strong"*), goal-oriented statements (e.g., *"I will complete this task successfully"*), and coping strategies (e.g., *"I can handle whatever comes my way"*). These scripts are not only used in sports psychology but also in clinical settings to manage stress and improve mental health [19]. For our system design, these self-talk scripts would serve as dedicated prompts for our dialog responses, tailored to different scenarios and emotional needs. By leveraging voice cloning technology, the system will present these generated contents in the user's deepfake voice, creating a more engaging and supportive self-talk experience.

### 2.2. System Implementation

In this subsection, we introduce the four core modules in our system and the technical details (Figure 1). The InnerSelf analyzes user's emotional state from the speech and implements appropriate dialog strategies to generate text responses. The text responses are then transformed into user's cloned voice, providing real-time conversational support and personalized suggestions for emotional well-being. When the dialogue is over, users can review their dialogue history and changes of emotions and thoughts, promoting more conscious self-regulation and modulation.

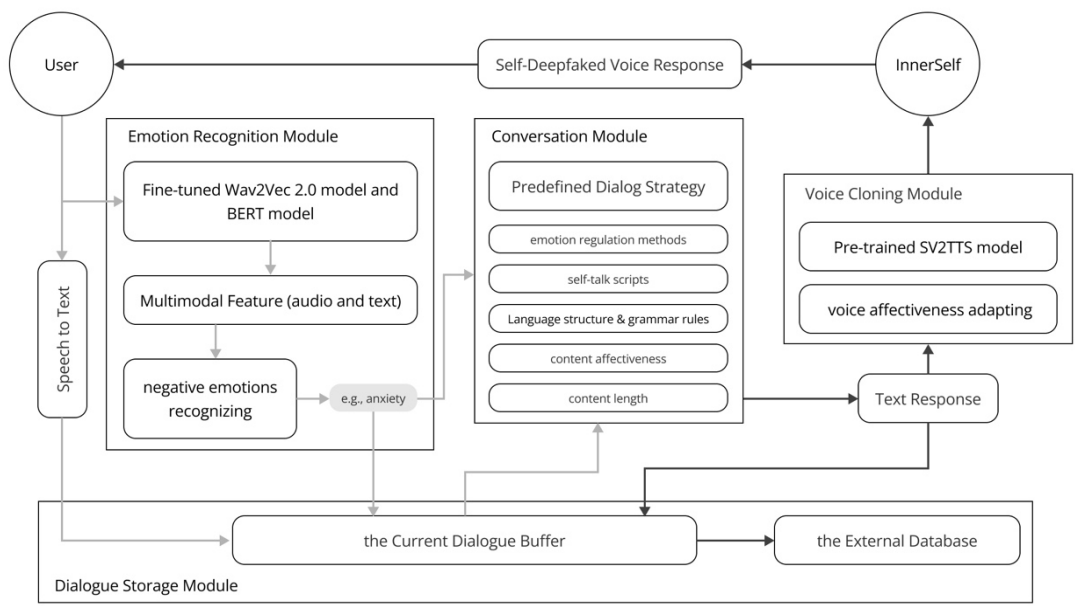

**Figure 1:** Technical Framework of InnerSelf

### 2.2.1. Emotion Recognition Module

As for emotion recognition, we utilize a Transformer architecture combined with multimodal fusion to analyze the user's emotion conveyed in their speech [20]. Leveraging the fine-tuned Wav2Vec 2.0 model [21] to handle the audio data and extract high-level features. We then apply a speech-to-text model to extract semantic features. The audio and text features are then fused to form a multimodal feature representation. Lately, a fully connected layer has been used to classify the fused features into emotional categories, with the Softmax function outputting the probability for each emotion category. Our approach performs well in recognizing negative emotions, including anxiety, sadness, shame or regret and anger.

### 2.2.2. Conversation Module

The user engages in a dialog with the InnerSelf, supported by the GPT-4 [22], as its built-in knowledge contains a wide range of emotional regulation techniques. Each time a new user's speech is received, conversation module would generate an appropriate response, which is then converted into a deepfake voice of the user. In user-InnerSelf interaction, this LLM-driven module selects the predefined dialog strategy according to the recognized emotions and the dialog context, which includes different emotion regulation methods and the self-talk scripts corresponding to each step, affectiveness and length of the content, and the language structure and grammar. Self-talk scripts include positive sentences or questions, depending on which emotion regulation method is implemented. The content affectiveness adhering to the principles of positive psychology, including the infusion of words or phrases of encouragement, empathy and understanding into the response to build rapport with users. The language structure and grammar, such as using the user's name, first person or non-first person pronouns to refer the user. Besides, for vocal responses, our system can adjust the pitch, volume, and speech rate according to the user's real-time emotions, resulting in speech that is both natural and empathetic to better encourage the user.

### 2.2.3. Voice Cloning Module

With the development of deep learning, neural networks are widely applied to the field of speech synthesis. Notable models such as Tacotron and WaveNet. Among them, SV2TTS model [23] can effectively capture and reproduce speaker characteristics through a pretrained speaker encoder, enabling effective training even with a small number of samples. Besides, this model supports data augmentation and transfer learning strategies, which further improves the model's performance and adaptability on new speech samples [23]. Thus, we selected the SV2TTS model to generate a naturally synthesized voice that closely resembles the user's voice, as user's deepfake voice.

For detailed steps: First, we preprocess user voice samples and proofread the speech transcripts to ensure the consistency of the text and audio content. Second, the samples converted to the cloned speech data format are used to fine-tune the pre-trained speaker encoder to adapt to the specific voice characteristics of the user. The speaker embeddings, along with text information, are then input into the synthesizer encoder, which further processes the data to generate Mel spectrograms. The WaveRNN vocoder converts these Mel spectrograms into clear and natural speech waveforms. We employ multiple rounds of training and feedback adjustments to optimize the model's performance, improving the similarity and naturalness of the synthesized voice. In addition, by collecting samples that cover a wide range of expressions and tonal variations and ensuring each sample is accurately labeled with its emotion, we train the model to precisely identify and emulate specific emotional vocal traits.

### 2.2.4. Dialogue Storage Module

Regarding dialogue storage, user's voice content and the changes of emotions and thoughts are recorded in both the current dialogue buffer and an external database. Initially, the recorded content is stored in a fixed-size buffer for the current talk, which includes the last $\alpha$ characters of data. We set $\alpha$ to 600 characters, a reasonable capacity to capture information from the last two minutes, providing sufficient conversation context for the system. As new dialogue inputs are received, the system processes data on a first-in, first-out (FIFO) basis, where new data is added to the end of the buffer, and old data exceeding the $\alpha$ threshold is removed from the beginning of the buffer. The removed data is then stored in chunks in the external database, ensuring data integrity and security, easy for future access and analysis.

## 3. Applications

The InnerSelf system designed for emotional well-being can be applied effectively across several areas. In the following section, we focus on three potential applications in detail.

### 3.1. Stop Negative Self-Talk Immediately

Negative self-talk is the voice inside that makes critical, negative or punitive comments. These voices are generally unconscious, uncontrolled speech that can limit one's ability to believe in themselves, and even increase the risk of mental health problems. To stop negative Self-talk immediately, a series of self-talk scripts can be applied to replace negative terms into positive ones (e.g., [24], [25]). After identifying negative emotions such as sadness, anger, anxiety, or self-deprecating comments in the user's speech, our system would help individuals reframe their negative talk into neutral or positive self-talk. If the user said, *"It's too difficult, I've tried everything."* the system can guide them to use affirmations to overcome anxiety, like *"Although I faced difficulties, I move forward, I learn from."* Or only replace absolute terms in a discourse, like *always* or *never*, with gentler terms, for example: turning *"I CAN'T EVER get things done on time. I'll NEVER be good at this."* to *"I OCCASIONALLY struggle with deadlines. I CAN get better at this."* The generated content would use the first person-singular pronouns with the deepfake self's voice. This process uses natural language processing techniques to ensure that the reworded message is contextualized.

### 3.2. Long-term Change in Thought Patterns

By providing personalized emotional intervention strategies to users, InnerSelf has the potential to address their negative thoughts in specific situations, contributing to both short-term and longterm emotional stability and improvement. Generally, Cognitive behavioral therapy requires close cooperation with therapists [26]. Besides, many emotion regulation methods focus on the benefits of reframing strategies, which instruct individuals to change their thought patterns to change their emotion [27]. In this context, our system allows users to exercise goal-directed self-talk, used with these reframing strategies by themselves. For instance, InnerSelf can guide users to identify and challenge negative thoughts using the cognitive restructuring method [28]. By employing Socratic questioning, users can explore and evaluate their negative assumptions step-by-step, find more reasonable ways of thinking, or develop personalized action plans.

During this process, our LLM-driven system would analyze cognitive change through change in the user's verbalizations, to generate responses. If the user feels anxious due to work pressure, InnerSelf can suggest specific stress management techniques and assist in creating an action plan. The system tracks the plan and provides necessary reminders and feedback. This continuous support helps users maintain positive changes and gradually improve their emotional state.

### 3.3. Psychological Assistance

Another opportunity for deploying the InnerSelf system is to use Innerself avatar therapist or to assist therapists for patients, particularly for those with mental health issues, such as mild to severe depression, post-traumatic stress disorder and anxiety disorders. Its potential benefits involve tracking and recording patients' emotional responses and changes through interactive dialogue, as well as identifying any related emotional triggers. It is crucial to obtain authentic these patients' health data outside regular clinical visits. With privacy consent issues resolved, therapists could access patients-generated health data, including daily mood fluctuations, levels of anxiety or depression, and any related emotional triggers, enabling more informed and targeted therapeutic interventions.

Additionally, inspired by the prior work [7], our system can also use the patient's self-deepfake voice, with its high emotional arousal and motivational reminders, to lead them through daily mindfulness meditation or other therapeutic exercises. We believe this stage-by-stage training helps patients integrate psychotherapy techniques into their daily activities, thereby enhancing the overall effectiveness of the therapy.

## 4. Conclusion and Future Work

This paper explores the psychological implications of hearing and talking to oneself and its application in personalized voice interfaces for emotional well-being. We present the InnerSelf, a voice interactive system that allows users to interact with their own cloned voice. We describe the system's technical framework and usage process and explore potential applications of the system for mental health support and emotion regulation. By fostering positive self-talk, our system aims to promote self-disclosure and regulation, reframe negative thoughts, and improve mental health. We plan to conduct a series of user studies to evaluate our system and identify novel design parameters for using self-deepfake voice in emotional regulation.

## Acknowledgements

This research was supported by Zhejiang University. We extend our special thanks to the researchers who assisted us throughout this process and the reviewers who provided us with constructive suggestions for improvement; their expertise and generous support were crucial to the successful completion of this research.

## Reference

[1] C. Rosa, M. Lassonde, C. Pinard, J. P. Keenan, and P. Belin, "Investigations of hemispheric specialization of self-voice recognition," *Brain Cogn.*, vol. 68, no. 2, pp. 204–214, 2008.

[2] N. A. Eger and E. Reinisch, "The impact of one's own voice and production skills on word recognition in a second language," *J. Exp. Psychol. Learn. Mem. Cogn.*, vol. 45, no. 3, pp. 552–571, 2019, doi: 10.1037/xlm0000599.

[3] H. Mitterer, N. A. Eger, and E. Reinisch, "My English sounds better than yours: Second-language learners perceive their own accent as better than that of their peers," *PLOS ONE*, vol. 15, no. 2, p. e0227643, Feb. 2020, doi: 10.1371/journal.pone.0227643.

[4] S. Ding *et al.*, "Golden speaker builder–An interactive tool for pronunciation training," *Speech Commun.*, vol. 115, pp. 51–66, 2019.

[5] R. M. Duncan and J. A. Cheyne, "Incidence and functions of self-reported private speech in young adults: A self-verbalization questionnaire," *Can. J. Behav. Sci. Rev. Can. Sci. Comport.*, vol. 31, no. 2, pp. 133–136, 1999, doi: 10.1037/h0087081.

[6] P. Sanches *et al.*, "HCI and Affective Health: Taking stock of a decade of studies and charting future research directions," in *Proceedings of the 2019 CHI Conference on Human Factors in Computing Systems*, in CHI '19. New York, NY, USA: Association for Computing Machinery, May 2019, pp. 1–17. doi: 10.1145/3290605.3300475.

[7] J. Kim and H. Song, "My Voice as a Daily Reminder: Self-Voice Alarm for Daily Goal Achievement," in *Proceedings of the CHI Conference on Human Factors in Computing Systems*, in CHI '24. New York, NY, USA: Association for Computing Machinery, May 2024, pp. 1–16. doi: 10.1145/3613904.3641932.

[8] C.-J. Chang and W.-C. Chien, "Towards a Positive Thinking About Deepfakes: Evaluating the Experience of Deepfake Voices in the Emotional and Rational Scenarios," in *Human-Computer Interaction*, M. Kurosu and A. Hashizume, Eds., Cham: Springer Nature Switzerland, 2024, pp. 311–325. doi: 10.1007/978-3-031-60405-8_20.

[9] Q. Xie *et al.*, "The Multi-Speaker Multi-Style Voice Cloning Challenge 2021," in *ICASSP 2021 - 2021 IEEE International Conference on Acoustics, Speech and Signal Processing (ICASSP)*, Jun. 2021, pp. 8613–8617. doi: 10.1109/ICASSP39728.2021.9414001.

[10] Y.-C. Lee, N. Yamashita, and Y. Huang, "Designing a Chatbot as a Mediator for Promoting Deep Self-Disclosure to a Real Mental Health Professional," *Proc. ACM Hum.-Comput. Interact.*, vol. 4, no. CSCW1, p. 31:1-31:27, May 2020, doi: 10.1145/3392836.

[11] Y.-C. Lee, N. Yamashita, Y. Huang, and W. Fu, "'I Hear You, I Feel You': Encouraging Deep Self-disclosure through a Chatbot," in *Proceedings of the 2020 CHI Conference on Human Factors in Computing Systems*, in CHI '20. New York, NY, USA: Association for Computing Machinery, Apr. 2020, pp. 1–12. doi: 10.1145/3313831.3376175.

[12] T. Conde, Ó. F. Gonçalves, and A. P. Pinheiro, "Stimulus complexity matters when you hear your own voice: Attention effects on self-generated voice processing," *Int. J. Psychophysiol.*, vol. 133, pp. 66–78, Nov. 2018, doi: 10.1016/j.ijpsycho.2018.08.007.

[13] A. P. Pinheiro, A. Farinha-Fernandes, M. S. Roberto, and S. A. Kotz, "Self-voice perception and its relationship with hallucination predisposition," *Cognit. Neuropsychiatry*, vol. 24, no. 4, pp. 237–255, Jul. 2019, doi: 10.1080/13546805.2019.1621159.

[14] T. B. Rogers, N. A. Kuiper, and W. S. Kirker, "Self-reference and the encoding of personal information," *J. Pers. Soc. Psychol.*, vol. 35, no. 9, pp. 677–688, 1977, doi: 10.1037/0022-3514.35.9.677.

[15] A. D. Argembeau, C. Comblain, and M. Van der Linden, "Affective valence and the self-reference effect: Influence of retrieval conditions," *Br. J. Psychol.*, vol. 96, no. 4, pp. 457–466, 2005, doi: 10.1348/000712605X53218.

[16] H. A. Sackeim and R. C. Gur, "Self-Deception, Self-Confrontation, and Consciousness," in *Consciousness and Self-Regulation: Advances in Research and Theory Volume 2*, G. E. Schwartz and D. Shapiro, Eds., Boston, MA: Springer US, 1978, pp. 139–197. doi: 10.1007/978-1-4684-2571-0_4.

[17] Z. Peng, Y. Wang, L. Meng, H. Liu, and Z. Hu, "One's own and similar voices are more attractive than other voices," *Aust. J. Psychol.*, vol. 71, no. 3, pp. 212–222, Sep. 2019, doi: 10.1111/ajpy.12235.

[18] P. Orepic, O. A. Kannape, N. Faivre, and O. Blanke, "Bone conduction facilitates self-other voice discrimination," *R. Soc. Open Sci.*, vol. 10, no. 2, p. 221561, Feb. 2023, doi: 10.1098/rsos.221561.

[19] A. T. Latinjak et al., "Self-Talk: An Interdisciplinary Review and Transdisciplinary Model," *Rev. Gen. Psychol.*, vol. 27, no. 4, pp. 355–386, Dec. 2023, doi: 10.1177/10892680231170263.

[20] K. Ezzameli and H. Mahersia, "Emotion recognition from unimodal to multimodal analysis: A review," *Inf. Fusion*, vol. 99, p. 101847, 2023.

[21] L. Pepino, P. Riera, and L. Ferrer, "Emotion Recognition from Speech Using Wav2vec 2.0 Embeddings," Apr. 08, 2021, *arXiv*: arXiv:2104.03502. doi: 10.48550/arXiv.2104.03502.

[22] OpenAI et al., "GPT-4 Technical Report," Mar. 04, 2024, *arXiv*: arXiv:2303.08774. doi: 10.48550/arXiv.2303.08774.

[23] Y. Jia et al., "Transfer Learning from Speaker Verification to Multispeaker Text-To-Speech Synthesis," in *Advances in Neural Information Processing Systems*, Curran Associates, Inc., 2018. Accessed: Jun. 18, 2024. [Online]. Available: https://proceedings.neurips.cc/paper_files/paper/2018/hash/6832a7b24bc06775d02b7406880b93fc-Abstract.html

[24] Kross, "Self-talk as a regulatory mechanism: how you do it matters.," *J. Pers. Soc. Psychol.*, vol. 106, no. 2, 2014, doi: 10.1037/a0035173.

[25] G. MAMASSIS and G. DOGANIS, "The Effects of a Mental Training Program on Juniors Pre-Competitive Anxiety, Self-Confidence, and Tennis Performance," *J. Appl. Sport Psychol.*, vol. 16, no. 2, pp. 118–137, Apr. 2004, doi: 10.1080/10413200490437903.

[26] A. C. Butler, J. E. Chapman, E. M. Forman, and A. T. Beck, "The empirical status of cognitive-behavioral therapy: A review of meta-analyses," *Clin. Psychol. Rev.*, vol. 26, no. 1, pp. 17–31, Jan. 2006, doi: 10.1016/j.cpr.2005.07.003.

[27] J. J. Gross, "Emotion Regulation: Current Status and Future Prospects," *Psychol. Inq.*, vol. 26, no. 1, pp. 1–26, Jan. 2015, doi: 10.1080/1047840X.2014.940781.

[28] A. Larsson, N. Hooper, L. A. Osborne, P. Bennett, and L. McHugh, "Using Brief Cognitive Restructuring and Cognitive Defusion Techniques to Cope With Negative Thoughts," *Behav. Modif.*, vol. 40, no. 3, pp. 452–482, May 2016, doi: 10.1177/0145445515621488.